\documentclass[11pt]{article}

\usepackage{amssymb}
\usepackage{amsthm}
\usepackage{amsmath}
\usepackage{graphicx}
\usepackage{psfrag}
\usepackage{subfigure}

\newcommand{\textfrac}[2]{{\textstyle{\frac{#1}{#2}}}}

\newcommand{\cE}{{\cal E}}

\def\bea{\begin{eqnarray}}
\def\eea{\end{eqnarray}}

\newcommand{\sfrac}[2]{{\textstyle{\frac{#1}{#2}}}}
\newcommand{\primelim}{\sideset{^\shortmid}{}\lim}

\theoremstyle{plain}
\newtheorem{theorem}{Theorem}[section]
\newtheorem{corollary}[theorem]{Corollary}
\newtheorem{lemma}[theorem]{Lemma}
\newtheorem{proposition}[theorem]{Proposition}
\newtheorem*{definition}{Definition}

\theoremstyle{remark}

\newtheorem*{remark}{Remark}

\begin{document}
\title{\sc Self-gravitating stationary spherically symmetric systems in relativistic galactic dynamics}

\author{\sc
Mikael Fj{\"a}llborg$^{1}$\thanks{Electronic address: 
{\tt mikael.fjallborg@kau.se}}, \
J.\ Mark
Heinzle$^{2}$\thanks{Electronic address: {\tt
mark.heinzle@univie.ac.at}}\,, \ and \ Claes
Uggla$^{3}$\thanks{Electronic
address: {\tt Claes.Uggla@kau.se}}\\
$^{1}${\small\em Department of Mathematics, University of Karlstad}\\
{\small\em S-651 88 Karlstad, Sweden} \\
{\small\em and }\\
{\small\em Department of Mathematics, Chalmers University of Tecnology}\\
{\small\em S-412 96 G\"oteborg, Sweden}  \\
$^{2}${\small\em Institute for Theoretical Physics, University of Vienna}\\
{\small\em Boltzmanngasse 5, A-1090 Vienna, Austria} \\
$^{3}${\small\em Department of Physics, University of Karlstad}\\
{\small\em S-651 88 Karlstad, Sweden}}

\date{}
\maketitle
\begin{abstract}
We study equilibrium states in relativistic galactic dynamics which
are described by stationary
solutions of the Einstein-Vlasov system for collisionless matter.
We recast the equations 
into a regular
three-dimensional system of autonomous first order ordinary
differential equations on a bounded state space. Based on a
dynamical systems analysis we derive new theorems that
guarantee that the steady state solutions have finite radii and
masses.
\end{abstract}
\centerline{\bigskip\noindent PACS number(s): 02.90.+p, 04.40.-b, 98.20.-d.} \vfill
\newpage

\section{Introduction}
\label{introduction}

The Einstein-Vlasov system describes a collisionless gas of
particles that only interact via the smoothed relativistic
gravitational field they generate collectively through their
averaged stress-energy. In this article we study the equilibrium
states of the spherically symmetric Einstein-Vlasov equations. This
topic is of considerable physical and mathematical interest:
Fackerell, Ipser, and Thorne have used these
models as models for relativistic star
clusters~\cite{fac68,ipstho68,ips69a,ips69b,fac70,fac71} (see also
these references for references to earlier literature);
Martin-Garcia and Gundlach studied these models in the context of critical
collapse~\cite{margun02}; Rein and Rendall approached the area from
a more mathematical point of view~\cite{reiren00} --- 
this paper will be taken as the starting point for the present
work.

It is straightforward to work with particles with different masses,
see e.g.~\cite{fac70,grav73}, but for simplicity we will restrict
ourselves to particles with a single mass, $m_0 \geq 0$, which
yields the same mathematical problem. A collisionless gas is
characterized by a non-negative phase space distribution
$f(x^\mu,p^\mu)$ defined on the future mass shell $PM$ of the
tangent space $TM$ of the spacetime, i.e., $x^\mu$ are spacetime
coordinates and $p^\mu$ are local coordinates of the four-momentum
with respect to the coordinate basis $x^\mu$ such that $p_\mu p^\mu
= -m_0^2$, $p^0>0$, where $p_\mu = g_{\mu\nu}p^\nu$, and where
$g_{\mu\nu}$ is the spacetime metric with signature $(-,+,+,+)$; 
throughout this article we set $c=1=G$, where $c$ is the speed of
light and $G$ is the gravitational constant.

For static spherically symmetric models, the metric can be written
as
\begin{equation}
\label{ds2} ds^2 = -e^{2\mu(r)}dt^2 + e^{2\lambda(r)}dr^2 +
r^2(d\theta^2 + \sin^2\theta d\phi^2)\:,
\end{equation}
and the Vlasov equation as
\begin{equation}
\frac{v}{\sqrt{1+v^{2}}}\cdot\nabla _{x}f-\sqrt{1+v^{2}}\mu
'\frac{x}{r}\cdot\nabla _{v}f=0\:,
\end{equation}
where $v$ is defined below. The non-negative energy-density and
radial- and tangential pressures are given by
\begin{subequations}\label{origrho}
\begin{align} \rho & =  \int f(r,v)\sqrt{1+v^{2}}dv\:,\\
p_{\rm rad}& =  \int f(r,v)\left(\frac{x\cdot
v}{r}\right)^{2}\frac{dv}{\sqrt{1+v^{2}}}\:,\\ p_{\rm T} & =
\sfrac{1}{2}\int f(r,v)\frac{|x\times
v|^{2}}{r^{2}}\frac{dv}{\sqrt{1+v^{2}}}\:.
\end{align}
\end{subequations}
The Vlasov equation admits a stationary solution given
by $f(E,L^2)$, where $E$ and $L$ are conserved quantities,
interpreted as particle energy and angular momentum, per unit mass;
they are defined by
\begin{equation}
\label{EL2} E:= e^{\mu(r)}\sqrt{1+v^2}\, ,\quad L^2 = r^2v_{\rm
T}^2\:,
\end{equation}
where we have followed~\cite{reiren00} and defined $v^2$ as
$v^2:=h_{\mu\nu}u^\mu u^\nu$, where $u^\mu = p^\mu/m_0$ and
$h_{\mu\nu}:=n_\mu n_\nu + g_{\mu\nu}$, where
$n_\mu=g_{\mu\nu}n^\nu$ and $n^\mu=e^{-\mu}(\partial_t)^\mu$;
$v_{\rm T}^2$ is defined as $v_{\rm T}^2=v^2-v_r^2$, where
$v_r=e^\lambda p^r/m_0$ and where $p^r$ is the radial coordinate
component of $p^\mu$. Note that the above objects are associated 
with the
spatial projection of the four-velocity: $v^\mu=h^\mu{}_\nu u^\nu$
and that they thus differ from the physical three-velocity, $V^\mu$,
which is defined by $u^\mu=\Gamma(n^\mu + V^\mu)$ and $n_\mu
V^\mu=0$, where $\Gamma=1/\sqrt{1-V^2}$; hence, e.g.,
$v^2=V^2/(1-V^2)$ and $E=\Gamma e^{\mu(r)}$.

Using that $f=f(E,L^2)$ yields:
\begin{subequations}\label{rhop}
\begin{align} \label{rhoE}\rho &= \frac{2\pi}{r} e^{-3 \mu}
\int\limits_{e^\mu}^{\infty} \int\limits_0^{L_{\mathrm{max}}^2}
f(E,L^2) \frac{E^2 dL^2
dE}{\sqrt{r^{2}(e^{-2\mu }E^2-1)-L^2}}\:, \\
\label{prE} p_{\rm rad} & = \frac{2\pi e^{-\mu }}{r^{3}}\int_{e^{\mu
}}^{\infty }\int_{0}^{L^{2}_{\rm max}}
f(E,L^{2})\sqrt{r^{2}(e^{-2\mu
}E^{2}-1)-L^{2}}dL^{2}dE\:,\\
p_{\rm T} & = \frac{\pi e^{-\mu }}{r^{3}}\int_{e^{\mu }}^{\infty
}\int_{0}^{L^{2}_{\rm max}}
f(E,L^{2})\frac{L^{2}dL^{2}dE}{\sqrt{r^{2}(e^{-2\mu
}E^{2}-1)-L^{2}}}\:,
\end{align}
\end{subequations}
where $L_{\mathrm{max}}^2 = r^2 (e^{-2 \mu} E^2-1)$.
%
%

Let $R \in (0,\infty]$ denote the radius of support of the system,
i.e., $\rho(r) = 0$ when $r \geq R$. We are interested in
gravitationally bound systems in equilibrium that either have finite
radii $R$ or possess a function $\rho$ that decreases 
sufficiently rapidly towards
infinity so that in both cases $\lim_{r\rightarrow R} \mu(r) =
\mu_R < \infty$. Because of the equilibrium assumption,
it follows that $v_r$ of the
individual particles has to satisfy $\lim_{r\rightarrow R} v_r = 0$.
We therefore require, since $E= e^{\mu(r)}\sqrt{1+v_r^2 + L^2/r^2}$,
that $f(E,L^2) = 0$ when $E \geq E_0 = e^{\mu_R}$, i.e., we assume
that there exists a cut-off energy that prevents particles from
escaping the gravitational field they create collectively.

The basic remaining equations of the Einstein-Vlasov system are given
by
\begin{subequations}\label{ein}
\begin{align}
\label{lambda}
  e^{-2\lambda}\,\left(2\frac{d\lambda}{d\xi} - 1\right) + 1 &= 8\pi\,r^2\,\rho , \\
\label{mu}   e^{-2\lambda}\,\left(2\frac{d\mu}{d\xi} - 1\right) - 1
&= 8\pi\,r^2\,p_{\rm rad}\:,
\end{align}
\end{subequations}
where $\xi:=\log r$. We define the mass function, $m(r)$, according to
\begin{equation}
e^{-2\lambda}:=1-\frac{2m(r)}{r}\:,
\end{equation}
which, together with~(\ref{lambda}), yields
\begin{equation}\label{meq}
\frac{dm}{d\xi} = 4\pi\,r^3\rho\:.
\end{equation}
%
Consequently, Eq.~(\ref{mu}) takes the form
\begin{equation}\label{mueq}
\frac{d\mu}{d\xi}=e^{2\lambda}\left( \frac{m}{r} + 4\pi\,r^2\,p_{\rm
rad}\right)\:.
\end{equation}
Since $f$ is assumed to be positive when $E<E_0$, it follows that
$\rho>0$ when $r<R$, and that $\mu$ is a monotonically increasing
function in $r$ (or $\xi =\log r$); it also follows that $0<m(r)\leq M:=m(R)$ when
$0<r\leq R$, where $M$ denotes the ADM mass.

In this paper we will consider distribution functions of the type
\begin{equation}
\label{D1} f= \phi(E)L^{2l}\:,
\end{equation}
where $\phi(E)$ is a non-negative function such that $\phi(E)=0$
when $E\geq E_0$; we also assume that $-1<l<\infty$.

Let us make the basic definitions
\begin{equation}
\eta := \ln(E_0)-\mu = \mu_R - \mu\:,\qquad \cE := E/E_0\:.
\end{equation}
Rewriting~\eqref{mueq} in terms of $\eta$ results in
\begin{equation}\label{etaeq}
\frac{d\eta}{d\xi}= -\left(1-\frac{2m}{r}\right)^{-1}\left(
\frac{m}{r} + 4\pi\,r^2\,p_{\rm rad}\right)\:.
\end{equation}
Note that $\eta$ is closely related to the redshift, $z$,
measured at the surface $R$: $z = e^\eta-1$.

For distributions of the type~(\ref{D1}) the density and the
radial pressure take the following form:
\begin{subequations}\label{prho}
\begin{align}
\label{p} p_{\rm rad} &= a\,r^{2l}\,e^{(2l+4)\,\eta
}\,g_{l+3/2}(\eta
)\:,\\
\label{rho} \rho &= (l+\sfrac{3}{2})\,a\,r^{2l}\,e^{(2l+4)\eta
}\,[2g_{l+3/2}(\eta ) + e^{-2\eta}g_{l+1/2}(\eta )]\:,
\end{align}
where the constant $a$ is given by
\begin{equation}
a=2^{l+3/2}\pi^{3/2}\,\frac{\Gamma(l+1)}{(l+3/2)\,\Gamma(l+3/2)}\:;
\end{equation}
here, $\Gamma(\cdot)$ denotes the Gamma function. The functions $g_i(\eta)$
are defined by
\begin{equation}\label{gidef}
 g_{i}(\eta):= \int_{e^{-\eta }}^{1}\, {\phi}(\cE )\,[\sfrac{1}{2}(\cE
 ^{2}-e^{-2\eta})]^{i}\,d\cE\:.
\end{equation}
\end{subequations}
where, with a slight abuse of notation, $\phi(E)=\phi(\cE)$; note
that $g_i>0$ and $\cE\,e^{\eta}= \sqrt{1+v^2} >1$. The
tangential pressure is given by $p_{\mathrm{T}}=(l + 1)p_{\rm rad}$,

Throughout the paper, the ratio $\sigma = p_{\mathrm{rad}}/\rho$ will play an
important role:
\begin{equation}\label{sig}
\sigma= \sigma(\eta):= \frac{p_{\rm rad}}{\rho }=
\frac{1}{3+2l}\,\left(\frac{g_{l+3/2}}{g_{l+3/2} +
\frac{1}{2}e^{-2\eta}g_{l+1/2}}\right)< \frac{1}{3+2l}\:.
\end{equation}
Note that this inequality is equivalent to $w<1/3$, where $w=p/\rho$
and where $p=(p_{\rm rad} + 2p_{\mathrm{T}})/3$ is the isotropic pressure; the
limit $w=1/3$, or equivalently $\sigma=1/(3+2l)$, corresponds to the
ultra-relativistic limit, i.e., radiation.

The equations~(\ref{meq}) and~(\ref{etaeq}) form the Einstein-Vlasov
system for equilibrium states; it is a two-dimensional
non-autonomous system (recall that $\xi=\log(r)$); the system is
closed via~(\ref{prho}) which provides the relation between 
$p_{\rm rad}$ and $\rho$ through the definition of $g_i(\eta)$ in
(\ref{gidef}). However, note that the system~(\ref{meq}) and
(\ref{etaeq}) is not just relevant for kinetic theory, but also for
any fluid with an anisotropic pressure $p_{\mathrm{T}}=(l + 1)p_{\rm rad}$ and
an equation of state that relates $p_{\rm rad}$ and $\rho$ (the
system reduces to one for an isotropic perfect fluid when $l=0$ so
that $p_{\mathrm{T}} = p_{\mathrm{rad}}$). The fundamental assumption in
kinetic theory that the distribution function be non-negative
automatically leads to natural physical conditions on $\rho$,
$p_{\mathrm{rad}}$, and $p_{\mathrm{T}}$; in particular, the density and
$p_{\mathrm{rad}}$ are positive when $r<R$. Since $l>-1$ it follows
that $p_{\mathrm{T}} > 0$ as well (the limiting case $l=-1$ describes a gas of
non-colliding particles with purely radial motion, or a fluid with
no tangential pressure). From a mathematical point of view one can
regard the kinetic case as a special anisotropic fluid case that
corresponds to a certain class of equations of state, implicitly
determined by (\ref{p}), (\ref{rho}) and (\ref{gidef}); however, recall that
$\sigma(\eta)<1/(3+2l)$ in the kinetic case.

If we assume the usual weak differentiability conditions so that
$\nabla_\mu\,T^{\mu\nu}=0$, where $T^{\mu\nu}$ is the
energy-momentum tensor, then we obtain the equation
\begin{equation}\label{pretolop}
\frac{d p_{\mathrm{rad}}}{d\xi}=2 l\,p_{\mathrm{rad}} +
(p_{\mathrm{rad}}+\rho)\frac{d\eta}{d\xi}\:,
\end{equation}
which, when (\ref{etaeq}) is inserted, yields the analogue of the
Tolman-Oppenheimer-Volkov equation. From~(\ref{pretolop}) it follows
that we can write $\sigma$ in the form
\begin{equation}\label{sigmaform}
\sigma = \left[ 3 + 2l + \frac{d}{d\eta} \log
g_{l+3/2}(\eta)\right]^{-1}\:.
\end{equation}

The outline of the paper is as follows. In Sec.~\ref{Dynsys} we
derive the dynamical system formulation: the Einstein-Vlasov
system is reformulated as an autonomous system of differential
equations on a bounded state space. This dynamical system is
subsequently analyzed in Sec.~\ref{dynanalys}. The results are
then 
interpreted in the standard physical variables 
to yield our main theorems
in Sec.~\ref{physinterpretation}; these theorems 
formulate various criteria that guarantee 
solutions of the Einstein-Vlasov system to have finite masses and
radii; hereby we generalize earlier work~\cite{reiren00}.
We conclude with a brief discussion in Sec.~\ref{concl}. In
Appendix~\ref{appdiff} we investigate various 
differentiability conditions that 
are needed in the paper and derive properties of the prominent function
$\sigma(\eta)$. In Appendix~\ref{appexist} we discuss existence and
uniqueness of solutions. In Appendix~\ref{appcenter} we perform a
center manifold analysis, 
which allows us
to sharpen one of the theorems when certain asymptotic
differentiability conditions are imposed. 
Finally, 
in Appendix D we present 
an extension of the monotonicity principle given in~\cite{waiell97}.

\section{Dynamical systems formulation}
\label{Dynsys}

In the following we reformulate the Einstein-Vlasov system as
an autonomous dynamical system on a bounded state space.
As a first step we introduce
dimensionless asymptotic homology invariant
variables~\cite{heietal03}
\begin{equation}\label{uvomega}
u = \frac{4\pi\,r^3\rho}{m}\, ,\qquad v =
\frac{\frac{m}{r}}{\sigma(1-\frac{2m}{r})};
\end{equation}
the variable $v$ is not to be confused with the velocity variable of
Section~\ref{introduction}. In the variables $(u,v,\eta)$ the
Einstein-Vlasov system as given by~\eqref{meq} and~\eqref{etaeq} takes
the form of an autonomous system,
\begin{subequations}\label{uvomeq}
\begin{align}
\label{du} \frac{du}{d\xi} &= u\,\left[3+2l - u -\left(1+\sigma -
{\sigma}'\right)\,h\right] \\
\label{dv} \frac{dv}{d\xi} &= v\,\left[(u-1)(1+2\sigma v) +
{\sigma}'\,h\right] \\
\label{deta} \frac{d\eta}{d\xi} &= -\sigma\,h\:,
\end{align}
where
\begin{equation}
h:=(1+\sigma u)v\:;
\end{equation}
\end{subequations}
note that henceforth $\sigma^\prime$ denotes $d\sigma/d\eta$.

In many cases, depending on the choice of the distribution function $\phi(\cE)$,
$\sigma^\prime(\eta)$ is a rather complicated function;
it is thus convenient to replace $\eta$
by a variable $\omega$ that is adapted to the choice of $\phi$,
so that $\sigma^\prime$ as a function of $\omega$ has a simpler form.
In addition, the degree of differentiability of $\sigma^\prime$ at $\eta =0$
can be improved; for instance, the function $\sigma^\prime = C_1 + C_2 \sqrt{\eta}$
is not $\mathcal{C}^1$ at $\eta = 0$; however, by choosing $\omega = \sqrt{\eta}$
we obtain $\sigma^\prime = C_1 +C_2 \omega$, which is smooth.

We introduce the variable $\omega$ by choosing $\omega(\eta)$ according
to the following rules:
$\omega(\eta)$ is smooth on $(0,\infty)$ and
strictly monotonically increasing, i.e., $d\omega/d\eta>0$;
$\omega(\eta)>0$ when $\eta>0$,
and $\omega \rightarrow 0$ when $\eta\rightarrow 0$.
Replacing $\eta$ by $\omega$ in the system~\eqref{uvomeq}
leads to
\begin{equation}
\frac{d\omega}{d\xi} =  -\sigma\,\frac{d\omega}{d\eta}\,h =
-\omega\,F\,h \qquad {\rm where} \qquad
F:=\sigma\,\frac{d\log\,\omega}{d\eta}\:;
\tag{$\ref{deta}^\prime$}
\end{equation}
we assume that $F = \sigma\, d\log \omega/d\eta$
is bounded for all $\eta$ and in the limit $\eta\rightarrow 0$.
The specific choice of $\omega$ depends on the type of distribution
function one is considering:
as we will see below, besides the trivial $\omega =\eta$, one useful choice is
\begin{equation}
\omega = k  \left( e^{b\eta} -1 \right)\:,
\end{equation}
where $k>0$ and $b>0$ are constants. In this case, $d\omega/d\eta = k b e^{b\eta}$
so that $d\omega/d\eta \rightarrow \mathrm{const}$ as $\eta \rightarrow 0$;
moreover, $F = b\sigma (1 +k/\omega)$, which is bounded in the limit
$\eta\rightarrow 0$ when $\sigma^\prime$ is bounded.
In general, it turns out that it
is often advantageous to choose $\omega$ to be of the
form
\begin{equation}\label{omegaeta}
\omega=k(\varphi(\eta )-1)\:,
\end{equation}
where $k$ is a positive constant and $\varphi(\eta)$ satisfies
$\varphi\geq 1$, $d\varphi/d\eta >0$, and $\varphi(0)=1$.

The last step is to introduce bounded variables $(U,V,\Omega)$
based on $(u,v,\omega)$:
\begin{equation}
U:=\frac{u}{1+u}\, \quad V:=\frac{v}{1+v}\, ,\quad
\Omega:=\frac{\omega}{1+\omega}\:;
\end{equation}
furthermore, let us
introduce a new independent variable $\lambda$ according to
\begin{equation}
\frac{d\lambda}{d\xi} = (1-U)^{-1}\,(1-V)^{-1}\:.
\end{equation}
This leads to the following dynamical system on the
bounded state space of the variables $(U,V,\Omega)$:
\begin{subequations}\label{dynsys}
\begin{align}
\label{dU}
  \frac{dU}{d\lambda } &= U(1-U)\,\left[[3 + 2l - (4+2l)U](1-V) -
  \left(1+\sigma - {\sigma}'\right)\,H\right], \\
\label{dV} \frac{dV}{d\lambda }&= V(1-V)\left[(2U-1)(1-V+2\sigma V)
+
{\sigma}'\,H\right], \\
\label{dOmega} \frac{d\Omega }{d\lambda }&=
-\Omega(1-\Omega)\,F\,H\:,
\end{align}
where
\begin{equation}
H:=(1-U+\sigma U)V\:;
\end{equation}
\end{subequations}
the quantities $\sigma, d\sigma/d\eta, F$ 
are now to be regarded as functions of $\Omega$.

The dynamical system~\eqref{dynsys} is polynomial in $U$ and $V$, so
that the system extends to the boundaries $U=0$, $U=1$, $V=0$,
$V=1$. The state space is thus given by $(U,V,\Omega) \in
[0,1]^2\times (0,1)$. The limit $\sigma \rightarrow 0$
corresponds to the Newtonian limit, since
the system~(\ref{dynsys}) then reduces to the system of equations
for the Newtonian case;
the Newtonian equations (for the polytropes) are also recovered 
for $\Omega\rightarrow 0$ (which is connected with 
$\sigma\rightarrow 0$, see~\eqref{sigmalim} below);
however, in order for that limit to exist,
$\lim_{\eta\rightarrow 0}\sigma^\prime$ must be defined. Hence,
before we can proceed to analyze the global dynamics of the
system~\eqref{dynsys} we must discuss the properties of
$\sigma^\prime$, in particular its degree of differentiability and
its limiting behavior as $\eta\rightarrow 0$.

Throughout the paper we will assume that $\sigma^\prime(\eta)$ is a
continuous function on $\eta\in (0,\infty)$, or, equivalently, that
$\sigma^\prime$ as a function of $\Omega$ is in
$\mathcal{C}^0(0,1)$. In Appendix~\ref{appdiff} we discuss in detail
which assumptions on the distribution function are required to
achieve a continuous function $\sigma^\prime$, here we merely state
the basic facts: we consider a distribution function such that
$\phi(\cE)$ satisfies: $\phi(\cE) = 0$ for $\cE>1$; $\phi \in
\mathcal{C}^0(0,1)$, and $\phi(\cE) \leq \mathrm{const} \: \cE^{-1 +
\epsilon}$ for some $\epsilon >0$ in a neighborhood of $\cE=1$. If
$-1<l<-1/2$ we assume, in addition, that $\phi$ is H\"older
continuous with index $\alpha> -l - 1/2$.

By assumption, $\sigma^\prime$ is continuous for $\eta>0$.
$\lim_{\eta\rightarrow 0} \sigma^\prime(\eta)$ need not exist;
however,
from now on we will assume that
$\sigma^\prime$
has upper and lower bounds in
the limit $\eta\rightarrow 0$, i.e.,
$k_1 < \sigma^\prime < k_2$
and hence $k_1\eta < \sigma < k_2\eta $
as $\eta\rightarrow 0$
(where $k_1, k_2$ are positive
constants).
A direct consequence is
\begin{equation}\label{sigmalim}
\lim_{\eta\rightarrow 0} \sigma(\eta) = 0\:.
\end{equation}
Higher differentiability of $\sigma^\prime$ requires
higher differentiability of $\phi(\cE)$.

In Appendix~\ref{appdiff} we discuss the `asymptotic polytropic case'.
For asymptotic polytropes we find in addition to~\eqref{sigmalim}
that
\begin{equation}\label{sigmaprime0}
\sigma^\prime_0:=
\lim_{\eta\rightarrow 0}\,{\sigma}'= \frac{1}{1+l+n}\:,
\end{equation}
where the constants $l$ and $n$ satisfy $-1<l<\infty$
and $1/2<n<\infty$
(the
latter condition is motivated from kinetic theory: $n>1/2$ implies
$n-3/2 >-1$, cf.~\eqref{dfdfd}, so that $g_i(\eta)$ exits).
The bounds on $l$ and $n$ imply that $0<\sigma^\prime_0<2$.
A statement equivalent to~\eqref{sigmaprime0} is that $\sigma^\prime$ is
in $\mathcal{C}^0[0,\infty)$
in the asymptotically polytropic case.
As we discuss in Appendix~\ref{appdiff},
the asymptotically polytropic case requires
a distribution function $\phi(\cE)$ with asymptotically
polytropic behavior, i.e.,
\begin{equation}\label{dfdfd}
\phi(\cE) = \phi_- (1-\cE)^{n-3/2} \, \left(1 + o(1) \right)
\end{equation}
as $\cE\rightarrow 1$ (with $n>1/2$); $\phi_- >0$ is a constant.

Let us conclude this section with an example: the relativistic
polytropes. This class of isotropic pressure models (i.e., $l=0$) is
characterized by the following equation of state:
\begin{equation}\label{relpol}
\rho = Kp^{1/\Gamma} + p/(\Gamma-1) \quad \Longleftrightarrow \quad
{\hat c}\rho = [x^{1/\Gamma} + x]/(\Gamma-1)\:,
\end{equation}
where $\Gamma, K$ are constants and ${\hat
c}=[(\Gamma-1)K]^{-\Gamma/(\Gamma-1)}$, $x={\hat c}p$.
In~\cite{heietal03} it was shown that
\begin{equation}
\omega = x^{(\Gamma-1)/\Gamma} = \sfrac{1}{\Gamma}\,(e^\eta - 1)\:,
\end{equation}
which is of the form $\omega=k(e^{b\eta}-1)$ with $k=1/\Gamma$ and
$b=1$, leads to
\begin{equation}
\sigma = (\Gamma - 1)\Omega\:, \quad
\frac{d\sigma}{d\eta}=\frac{(\Gamma-1)\Gamma\,e^\eta}{(e^\eta +
\Gamma -1)^2} = \sfrac{\Gamma-1}{\Gamma}\,((\Gamma-1)\Omega +
1)(1-\Omega)\:,
\end{equation}
and a dynamical system with purely polynomial right hand sides.

In~\cite{ips69b,fac70} the kinematic interpretation of the following
class of equations of state was developed: $\rho = Kp^{1/\Gamma_1} +
p/(\Gamma_2-1)$ (also denoted as relativistic polytropes in the
given reference); when $\Gamma_1=\Gamma_2=\Gamma$ we recover the
previous example. The associated distribution functions are given
in~\cite{ips69b}, but reasonably simple expressions are only
obtained for $\Gamma_1=\Gamma_2=\Gamma$ and for special values of
$\Gamma$. For instance, $\Gamma=5/4$ yields $\phi =
\phi_-\,\cE^{-5}\,(1-\cE^2)^{5/2}$.

\section{Dynamical systems analysis}
\label{dynanalys}

In Appendix~\ref{appexist} we establish (global) existence and
uniqueness of solutions of the dynamical system~\eqref{dynsys} on
the state space ${\mathbf X}=(U,V,\Omega) \in [0,1]^2\times (0,1)$,
where we assume that $\sigma^\prime$ be continuous on $(0,1)$. In
this section, we analyze the global dynamics of the flow of the
dynamical system on $\mathbf{X}$.

The key ingredient in the dynamical systems analysis is the fact
that $\Omega$ is strictly decreasing in the interior of the state
space ${\mathbf X}$, which entails that the $\alpha$- and
$\omega$-limit sets of orbits (e.g., fixed points) reside on the
boundaries of $\mathbf{X}$. Standard local dynamical systems
analysis requires the dynamical system to be at least
$\mathcal{C}^1$ --- in that case the system can be linearized at the
fixed points. The dynamical system~\eqref{dynsys} is $\mathcal{C}^1$
on the state space $[0,1]^2\times (0,1)$, if $\sigma^\prime$ is in
$\mathcal{C}^1(0,1)$.
However, we will not make this smoothness assumption on
$\sigma^\prime$; for the following we merely assume $\sigma^\prime$
to be in $\mathcal{C}^0(0,1)$; in our proofs, monotonicity arguments
will replace the standard arguments from local dynamical systems
analysis.

\begin{figure}[h]
\centering
\includegraphics[height=0.40\textwidth]{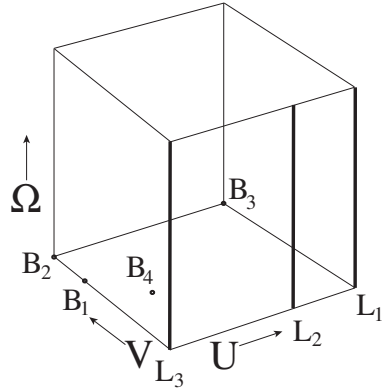}
\caption {The state space and fixed points; note that $B_4$ only
exists for certain ${\sigma}'_0$ (or equivalently $n$) values and
that an additional line of fixed points appears when ${\sigma}'_0=1$,
see Table~\ref{fixtab}.} \label{Statespace}
\end{figure}

The inclusion of the boundary $\Omega =0$ in the dynamical systems
analysis depends on the differentiability properties of
$\sigma^\prime$ at $\Omega =0$. If $\sigma^\prime$ is in
$\mathcal{C}^0[0,1)$ (instead of merely in $\mathcal{C}^0(0,1)$),
then the dynamical system continuously extends to the state space
$[0,1]^2\times [0,1)$; this is the case for asymptotically polytropic
distributions. If $\sigma^\prime$ (as a function of
$\Omega$) is in $\mathcal{C}^1[0,1)$, then the dynamical system
smoothly extends to $[0,1]^2\times [0,1)$. Although we will not make
these assumptions in the following, it is useful to consider the
case of smooth $\sigma^\prime$ to obtain a more intuitive feeling
for the properties of the flow of the dynamical system.
Figure~\ref{Statespace} depicts the state space $\mathbf{X}$ and its
extension to $\Omega =0$; for figures describing the dynamics on
$\Omega=0$ when $l=0$, see the appendix in~\cite{heiugg03}.
Table~\ref{fixtab} lists the fixed points and the eigenvalues of the
linearizations of the dynamical system.

\begin{table}[ht]
\begin{center}
\begin{tabular}{|c|c|c|c|c|c|}\hline
Fix points & $U$ & $V$ & $\Omega$ & Eigenvalues & Restrictions\\
\hline
$L_1$ & 1 & 0 & $\Omega _0$ & 1; 1; 0 &\\
$L_2$ & $\frac{3+2l}{4+2l}$ & 0 & $\Omega _0$
& $-\frac{3+2l}{4+2l}$; $\frac{1+l}{2+l}$; 0 & \\
$L_3$ & 0 & 0 & $\Omega _0$ & -1; $3+2l$; 0 & \\
$L_4$ & $U_0$ & 1 & 0 & $-(1-U_0)$; $\lambda_3$; 0 & ${\sigma}'_0=1$\\
$B_1$ & 0 & $\frac{1}{1+{\sigma}'_0}$ & 0 & $\frac{2(2+l){\sigma}'_0
- 1}{1+{\sigma}'_0}$;
$\frac{{\sigma}'_0}{1+{\sigma}'_0}$; $\frac{-F_0}{2+{\sigma}'_0}$ & \\
$B_2$ & 0 & 1 & 0 & $-(1-{\sigma}'_0)$; $-{\sigma}'_0$; $-F_0$ & \\
$B_3$ & 1 & 1 & 0 & 0; 0; 0 & \\
$B_4$ & $\frac{3+l-n}{2(2-n)}$ & $\frac{2(1+l)}{3+2l-2{\sigma}'_0}$
& 0 & $\alpha[1 \pm (5+3l-n)\sqrt{b}]$;$\lambda_3$ & $3+l-n>0$
\\ \hline
\end{tabular}
\end{center}
\caption{Fixed points and associated eigenvalues; the third
eigenvalue $\lambda_3$ always has the same form for all fixed points
on $\Omega=0$, namely $\lambda_3=-F_0V(1-U)$, $F_0=F(0)$; $\alpha =
\frac{(1+l)(n-2)}{2V(n+l-1)^2}$, $b=1+22l+33l^2+8l^3+2(11+13l)n -
(7+8l)n^2$, where $U$ and $V$ take the relevant fixed point values.}
\label{fixtab}
\end{table}

In the following we investigate the global dynamics of the dynamical
system on the state space $[0,1]^2\times(0,1)$. The most useful tool
for our analysis will be the monotonicity principle,
see~Appendix~\ref{appmon}.

\begin{lemma}\label{omegalimit}
The $\omega$-limit of every interior orbit is located on $\Omega =0$.
\end{lemma}
\begin{proof}
From equation~\eqref{dOmega} we see that $\Omega$
is a strictly monotonically decreasing function on the state
space $[0,1]^2 \times (0,1)$ except for on the side face $V=0$,
where $\Omega = \mathrm{const}$.
It follows from the monotonicity principle that the
$\omega$-limit of every interior orbit is located
either on $\Omega=0$ or $V=0$.
We prove the lemma by excluding the latter possibility:

Suppose that there exists an orbit $\gamma$ that possesses an
$\omega$-limit point on $V=0$ (with $U\neq 0$ and $U\neq 1$).
Then a point $P$ on $L_2$ must be in $\omega(\gamma)$ as well,
as follows from considering the structure of the flow on the
side face $V=0$.
Since $d V/d\lambda > 0$ in a neighborhood $\mathcal{U}$
of $L_2$, the orbit $\gamma$
cannot remain in $\mathcal{U}$, but must leave and re-enter
$\mathcal{U}$ infinitely many times.
The boundedness of the r.h.s.\ of the dynamical system implies that
the return time is bounded below and that $\Omega$ must decrease
by at least $\Delta\Omega > 0$ during that time.
However, this is in contradiction with the fact
that $P$ is an $\omega$-limit of $\gamma$.
Using similar (simpler) arguments we can also exclude that $L_1$ and $L_3$
are eligible $\omega$-limits. Therefore, the side face $V=0$
does not contain $\omega$-limit points of any interior orbit,
which leaves $\Omega =0$ as the only attractor.
\end{proof}
\begin{remark}
The proof of Lemma~\ref{omegalimit}
does not require the assumption that the
r.h.s.\ of the dynamical system be
$\mathcal{C}^1$ on $[0,1]^2\times(0,1)$ --- we have merely assumed
continuity of $\sigma^\prime$ on $(0,1)$.
However, if $\mathcal{C}^1$ is assumed, the line of argument
can be simplified by noting that $L_1$, $L_2$, $L_3$ consist
of transversally hyperbolic fixed points: points on
$L_1$ are sources, points on $L_2$ and $L_3$ are saddles.
\end{remark}

\begin{lemma}
The $\alpha$-limit set of an interior orbit is either located on
$L_1$, $L_2$, or $\Omega =1$.
A two-parameter set of orbits originates from $L_1$, a one-parameter set
from $L_2$.
\end{lemma}

\begin{proof}
The monotonicity of $\Omega$ implies that the $\alpha$-limit of an
interior orbit must lie either on $\Omega = 1$ or $V=0$. Points on
$L_3$ cannot acts as $\alpha$-limits: the line of argument resembles
the proof of Lemma~\ref{omegalimit}, where we use that $dU/d\lambda
> 0$ and $d V/d\lambda <0$ in a neighborhood of $L_3$. In a
neighborhood of $L_1$ we observe that $dU/d\lambda <0$ and $d
V/d\lambda > 0$; hence, $L_1$ is a source for a two-parameter family
of orbits. Orbits originating from $L_2$ correspond to regular
solutions of the Einstein-Vlasov equations, as we will see in
Section~\ref{physinterpretation}. The existence of a one-parameter
set of such solutions has been proved in~\cite{Baumgarte/Rendall}.
\end{proof}

\begin{proposition}\label{n3prop}
Suppose there exists $\hat{\eta} > 0$, arbitrarily small, such that
$\sigma(\eta)$ satisfies
\begin{equation}\label{Zcondi}
\sigma^\prime \geq \frac{1}{4 + 2 l} + \frac{7 + 4 l}{4 + 2 l} \:\sigma
\end{equation}
for all $\eta \leq \hat{\eta}$. Then the $\omega$-limit of every interior orbit lies
on $(\Omega = 0) \cap (V=1)$.
\end{proposition}

\begin{proof}
Let $\hat{\Omega}$ be the value of $\Omega$ associated with $\hat{\eta}$.
As the first step in the proof we show that the function $Z$ defined by
\begin{equation}\label{Zfctdef}
Z:= \frac{U}{1-U} \Big(\frac{V}{1-V}\Big)^{(3+2l)}
\end{equation}
is strictly monotonically
increasing on $(0,1)^2 \times (0,\hat{\Omega})$ if and only if
condition~\eqref{Zcondi} holds.

A straightforward computation leads to
\begin{equation}\label{Zdot}
Z^{-1}\: \frac{d Z}{d\lambda} = [2(2+l)\sigma^\prime - 1 - (7+4l) \sigma]
[V(1-U) + \sigma U V] + A \geq A > 0\:;
\end{equation}
here, we have used that
$A:=U[2(1+l)(1-V) + 2 (3 +2 l) V \sigma (1+ \sigma)]>0$ and the
assumption~\eqref{Zcondi}.
Conversely, if
$d Z/d\lambda >0$
for all $(U,V) \in (0,1)^2$ (and $\Omega< \hat{\Omega}$),
then~\eqref{Zcondi} follows by letting $U\rightarrow 0$
in~\eqref{Zdot}.


On the side faces $\{(U,V)\,|\,U=0, V <1\}$ and $\{(U,V)\,|\,V=0, U <1\}$
the function $Z$ assumes its infimum: $Z=0$.
The monotonicity principle, cf.~Appendix~\ref{appmon}, states
that $\omega$-limits cannot be located on this set.
As a consequence, the $\omega$-limit of every orbit of the considered
set $(0,1)^2 \times (0,\hat{\Omega})$
must be contained on $(U=1)$ or $(V=1)$.
Since $\Omega\rightarrow 0$ for all interior orbits,
we find that the $\omega$-limit of every orbit in $(0,1)^3$ lies
on $(\Omega = 0) \cap (U=1)$ or $(\Omega = 0) \cap (V=1)$ (or the union).
It remains to prove that the set $(\Omega =0) \cap (V=1)$ is the exclusive attractor.

Consider Eq.~\eqref{dV}: $dV/d\lambda \propto [(2 U -1) (1 -V + 2
\sigma V) + \sigma^\prime H]$. In the case $U \geq 1/2$, $d
V/d\lambda$ is manifestly positive. In the case $U < 1/2$,
positivity is guaranteed under the condition
\begin{equation}\label{Vprimeposcond}
\sigma^\prime > \frac{1 - 2 U}{1-U +\sigma U} \frac{1-V+2 \sigma V}{V}\:.
\end{equation}
Since
$(1-2U)/(1-U+\sigma U) \leq 1$ for all $U\in[0,1/2)$, a sufficient
condition for~\eqref{Vprimeposcond} to hold is
%
$\sigma^\prime > v^{-1} + 2 \sigma$,
%
which, by assumption~\eqref{Zcondi}, is satisfied if $v > 2(2+l)
(1-\sigma)^{-1}$. Therefore, there exists an $\epsilon> 0$ such
that $d V/d\lambda$ is positive on $\mathcal{S} := [0,1] \times
(1-\epsilon,1)\times (0,\epsilon)$. Consequently, $\mathcal{S}$ is a
future invariant set of the dynamical system and $V$ is a monotonic
function on $\mathcal{S}$; by the monotonicity principle, the
$\omega$-limit of every orbit in $\mathcal{S}$ must lie on $(\Omega
=0) \cap (V=1)$.

Now suppose that there exists an orbit $\gamma$ whose $\omega$-limit
$\omega(\gamma)$ contains a point on the set $(\Omega = 0) \cap
(U=1)$ (with $V \neq 1$). The flow on this set is known explicitly;
it follows that the point $(1,1,0)$ is in $\omega(\gamma)$.
Consequently, for all $\varepsilon >0$ there exists
$\lambda_\varepsilon$ so that $V(\lambda_\varepsilon) \geq  1
-\varepsilon$ and $\Omega(\lambda_\varepsilon) < \varepsilon$. For
sufficiently small $\varepsilon$ we find
$\gamma(\lambda_\varepsilon) \in \mathcal{S}$, whereby
$\gamma(\lambda) \in \mathcal{S}$ $\forall \lambda \geq
\lambda_\varepsilon$, since $\mathcal{S}$ is future invariant, and
consequently $V(\lambda)\rightarrow 1$ as $\lambda\rightarrow
\infty$. This is a contradiction to the assumption about the
$\omega$-limit of $\gamma$. Accordingly, the $\omega$-limit of every
orbit must lie on $(\Omega = 0) \cap (V=1)$.
\end{proof}

We now consider the surface
\begin{equation}
P(u,v) := 2 u + v - 2(3 + 2l) = 0
\end{equation}
in the state space of the variables $(u,v,\omega)$; equivalently, $P
= 0$ generates the surface $[4(2+ l) - (9 + 4l)V]U + (7 + 4 l) V
-2(3 + 2l)= 0$ in the state space $\mathbf{X} = \{(U,V,\Omega) \in
[0,1]^2\times (0,1)\}$. However, since $u$ and $v$ are finite on $P
= 0$ and the associated equations are simpler, it is preferable to
perform the following analysis in the variables $(u,v,\omega)$.

\begin{lemma}\label{5lemma}
Suppose that there exists $\check{\eta}>0$, such that $\sigma(\eta)$ satisfies
\begin{equation}\label{memcondi}
\sigma^\prime \geq \frac{1}{2(3 + 2l)} + k \:\sigma
\end{equation}
for all $\eta \leq \check{\eta}$, where $k = 3 + 2 l + \frac{1}{2}$
if $l \geq -\frac{3}{4}$, and $k =2$ if $ -1 < l < -\frac{3}{4}$.
Then
\begin{equation}\label{memprop}
\frac{d P}{d \xi}\Big|_{P=0} \geq 0\:,
\end{equation}
when $\omega \leq \check{\omega}$ ($u\neq 0$, $v\neq 0$); here
$\check{\omega}$ is the value of $\omega$ associated with
$\check{\eta}$.
\end{lemma}

\begin{proof}
Using the system~(\ref{uvomeq}) we obtain by a straightforward calculation
\begin{equation}
\frac{d P}{d\xi} \Big|_{P=0} =
v \:\Big[ -1 + 2 s \sigma^\prime - 2 \sigma^2 u^2 +
\sigma \left( 2 (1 - 3 u)  u +
2 s (-2 + (2 + \sigma^\prime) u) \right)\, \Big]\:,
\end{equation}
where $s = 3 + 2l$. It follows that
\begin{equation}\label{iffcondi}
\frac{d P}{d\xi} \Big|_{P=0} > 0 \quad\Leftrightarrow\quad
\sigma^\prime \geq \frac{1}{2 s} + \tilde{k}(u,\eta) \sigma\:,
\end{equation}
where
\begin{equation}
\tilde{k}(u,\eta) = \frac{-4 s (u-1) + u(-3 +2 (3+\sigma) u)}{2 s (1 + \sigma u)}\:.
\end{equation}
Assume that $ l > -\frac{3}{4}$, so that $s > \frac{3}{2}$, and
define $k = s + \frac{1}{2}$. We have
\begin{equation}\label{ktildek}
2 s (1 +\sigma u ) \Big[ \tilde{k}(u,\eta) - k\Big] = -(s-u)[2s - 3
+ 2u(3+\sigma)] - s(2s-1)\sigma u< 0\:,
\end{equation}
where we have used that $0<u<s$ on $P=0$. Accordingly, $k >
\tilde{k}(u,\eta)$ on $P=0$;
therefore, from~\eqref{iffcondi}, we conclude
that~\eqref{memcondi} is a sufficient condition for~\eqref{memprop}
to hold. Incidentally, note that in~\eqref{ktildek} equality holds
in the limit $\eta\rightarrow 0$ ($\sigma\rightarrow 0$) and
$u\rightarrow s$; hence, $k = s + \frac{1}{2}$ is the smallest
constant such that~\eqref{memcondi} implies~\eqref{memprop}. The
proof in the case $l < -\frac{3}{4}$ is analogous.
\end{proof}

\begin{proposition}\label{n5prop}
Suppose that there exists $\check{\eta}>0$, such that $\sigma(\eta)$ satisfies
\begin{equation}\label{n5condi}
\sigma^\prime \geq \frac{1}{2 (3 + 2l)} + k \:\sigma
\end{equation}
for all $\eta \leq \check{\eta}$, where $k = 3 + 2 l + \frac{1}{2}$
if $l \geq -\frac{3}{4}$, and $k =2$ if $ -1 < l < -\frac{3}{4}$.
Consider an orbit $\gamma$ whose $\alpha$-limit is a point on $L_1$
or $L_2$ with $\Omega \leq \check{\Omega}$ ($\eta \leq
\check{\eta}$). Then the $\omega$-limit of $\gamma$ is located on
$(\Omega = 0) \cap (V=1)$.
\end{proposition}

\begin{proof}
Consider the region $\mathcal{P}$ of the state space defined by $P >
0$ and $\Omega \leq \check{\Omega}$; this set is future invariant by
Lemma~\ref{5lemma}. On the set $\mathcal{P}$, $V$ is a strictly
monotonically increasing function. To see this, recall that
$dV/d\lambda$ is manifestly positive when $U\geq 1/2$, and from
manipulating~\eqref{Vprimeposcond} that
\begin{equation}\label{Vprimeposcondiagain}
\sigma^\prime > \frac{1-u}{1+\sigma u} v^{-1} + 2 \sigma
\end{equation}
is a sufficient condition for $d V/d\lambda > 0$ on $U \in [0,1/2)$ ($V\neq 1$).
By assumption we have
$\sigma^\prime > \frac{1}{2(3+2l)} + k \sigma \geq \frac{1}{2(3+2l)}
+ 2 \sigma$\:;
therefore, condition~\eqref{Vprimeposcondiagain} is satisfied if
$(1-u)^{-1} (1+\sigma u) v \geq 2(3+2l)$.
It is straightforward to show that this relation holds for $(u,v)$ such that $P(u,v) =0$,
from which follows that it is satisfied for $(u,v)$ such that $P(u,v) > 0$ as well.
This establishes the claim that $V$ is monotonic on $\mathcal{P}$.

Since $\mathcal{P}$ is future invariant and $V$ is a monotonically
increasing function on $\mathcal{P}$, the monotonicity principle in
conjunction with Lemma~\ref{omegalimit} implies that the
$\omega$-limit of every orbit in $\mathcal{P}$ resides on $(\Omega=
0) \cap (V=1)$. Orbits that originate from $L_1$ satisfy $P >0$
initially, i.e., these orbits are contained in $\mathcal{P}$. For an
orbit $\gamma$ whose $\alpha$-limit is a point on $L_2$ we have $P
\rightarrow 0$ as $\lambda\rightarrow -\infty$.
However, a simple computation shows that the unstable
manifold of every point on $L_2$ (with $\Omega \leq \check{\Omega}$)
lies in $\mathcal{P}$. The orbit $\gamma$ thus lies in $\mathcal{P}$
for all $\lambda$ and must possess an $\omega$-limit on $(\Omega= 0)
\cap (V=1)$. This concludes the proof of the proposition.
\end{proof}

\begin{remark}
In the asymptotically polytropic case,
$\sigma^\prime_0 = \lim_{\eta\rightarrow 0} \sigma^\prime$ exists:
$\sigma_0^\prime = (1 + l +n)^{-1}$.
Therefore, in the asymptotically polytropic case,
the assumptions of Proposition~\ref{n3prop}, see~\eqref{Zcondi}, are satisfied if
\begin{equation}
n < 3 + l \:. \tag{\ref{Zcondi}${}^\prime$}
\end{equation}
Furthermore,
\begin{equation}\label{n5condiaga}
n < 5 + 3 l   \tag{\ref{n5condi}${}^\prime$}
\end{equation}
is sufficient for condition~\eqref{n5condi} in Proposition~\ref{n5prop} to hold:
namely, if $n < 5 + 3l$ ($\sigma_0^\prime > (6 + 4l)^{-1}$), then there exists
$\check{\eta}$ such that $\sigma^\prime > (6 + 4l)^{-1} + k \sigma$ for all
$\eta \leq \check{\eta}$.
\end{remark}

In the asymptotically polytropic case, $\sigma^\prime(\eta)$ is
continuous on $\eta\in[0,\infty)$. If $\sigma^\prime$ (as a function of $\Omega$)
is even $\mathcal{C}^1[0,1)$, then the dynamical system extends smoothly to
the state space $[0,1]^2\times [0,1)$. Smoothness of
$\sigma^\prime(\eta)$ requires smoothness assumptions on
$\phi(\cE)$, which we refrain from discussing here; let us merely
note that, e.g., polytropic functions $\phi(\cE) = \phi_-
(1-\cE)^{n-3/2}$, and linear combinations thereof, satisfies the
requirements. If $\sigma^\prime(\eta)$ is $\mathcal{C}^1$ (or,
more generally, if there exists a variable $\omega$ such that
$\sigma^\prime$ is $\mathcal{C}^1$ as a function of $\omega$), then
Proposition~\ref{n3prop} can be improved:

\begin{proposition}\label{nequals3prop}
Assume that there exists a variable $\omega$ such that $\sigma^\prime|_{\eta(\omega)}$
is $\mathcal{C}^1[0,\epsilon)$ (for some $\epsilon >0$).
If
\begin{equation}
\sigma_0^\prime \geq \frac{1}{4 + 2l}\:,\qquad \text{or, equivalently,}\qquad
n \leq 3 + l\:,
\end{equation}
then the $\omega$-limit of every interior orbit lies
on $(\Omega = 0) \cap (V=1)$.
\end{proposition}

\begin{remark}
Recall that $\Omega$ can be chosen to be a smooth function of $\omega$; hence
$\sigma^\prime$ is $\mathcal{C}^1$ as a function of $\omega$
if and only if it is $\mathcal{C}^1$ as a function of $\Omega$.
In the simplest case, $\sigma^\prime$ is $\mathcal{C}^1[0,\epsilon)$
as a function of $\eta$ and thus automatically satisfies the assumptions of the proposition.
\end{remark}

\begin{proof}
The proof of the proposition is based on center manifold theory; it is given
in Appendix~\ref{appcenter}.
\end{proof}

\section{Physical interpretation}
\label{physinterpretation}

We define a \textit{regular solution} of the static Einstein-Vlasov
equations as a solution that possesses a regular `potential' $\mu$.
As $r\rightarrow 0$, $\mu(r)$ converges to a central value
$\mu_{\mathrm{c}}$; equivalently, $\eta(r) \rightarrow
\eta_{\mathrm{c}}$ as $r\rightarrow 0$. This implies that the limits
$\lim_{r\rightarrow 0} (r^{-2l}\rho)$ and $\lim_{r\rightarrow 0}
(r^{-2l} p_{\mathrm{rad}})$ exist, as follows from~\eqref{prho}.
Accordingly, $\rho \rightarrow \infty$ as $r\rightarrow 0$, if
$l<0$; $\rho \rightarrow \mathrm{const}$, if $l=0$; $\rho
\rightarrow 0$, if $l>0$. Hence, regular solutions are characterized
by a regular metric; however, $\rho$ and $p_{\mathrm{rad}}$ are only
regular when $l\geq 0$.

For regular solutions, the variables $U$ and $V$ satisfy
$U\rightarrow (3+2l)/(4+2l)$ and $V\rightarrow 0$ as $r\rightarrow 0$;
this is a direct consequence of~\eqref{uvomega}
(where we use $dm/d r = 4\pi r^2 \rho$ and the behavior of $\rho$ as $r\rightarrow 0$).
In the dynamical systems formulation, regular solutions thus correspond
to orbits whose $\alpha$-limit is a point on the line $L_2$.

\begin{remark}
Orbits whose $\alpha$-limit is a point on $L_1$ correspond to solutions
with a negative mass singularity at the center.
For a discussion of such solutions (in the Newtonian case)
see~\cite{heiugg03}.
\end{remark}

\begin{lemma}\label{finite}
An orbit in the state space with $\omega$-limit on $(\Omega =0) \cap (V=1)$
corresponds to a solution of the Einstein-Vlasov equations
with finite radius and mass.
\end{lemma}

\begin{proof}
Integration of the defining equation for $\lambda$,
$d \log r/d\lambda = (1 - U) (1-V)$, leads to
\begin{equation}\label{rint}
r = r_0 \exp \left[ \int_{\lambda_0}^\lambda \big(1 - U(\lambda)\big)
\big(1 - V(\lambda)\big) d\lambda \right]\:,
\end{equation}
where $\lambda_0$ and $r_0$ are such that $(U,V,\Omega)(\lambda_0)$
corresponds to $(m(r_0), \eta(r_0), r_0)$. In order to prove that
the solution possesses a finite radius we show that $r\rightarrow R
< \infty$ as $\lambda \rightarrow \infty$.

Define $\delta U = 1-U$ and $\delta V = 1-V$.
Using these variables, Eq.~\eqref{dV} takes the form
\begin{equation}
- \frac{d \delta V}{d\lambda} = (\delta V) V
\left[ \delta U \big(\sigma^\prime (1-\sigma) V - 2 \delta V - 4\sigma V \big)
+ \delta V + \sigma V ( 2 +\sigma^\prime) \right]\:.
\end{equation}
By assumption, $\sigma \rightarrow 0$ and $V\rightarrow 1$ ($\delta V \rightarrow 0$)
as $\lambda \rightarrow \infty$;
therefore, for sufficiently large $\lambda$,
\begin{equation}\label{ddeltaV}
- \frac{d \delta V}{d\lambda}  \geq \delta U \delta V \sigma^\prime (1-\epsilon)
\end{equation}
where $\epsilon > 0$ is small.
Inserting~\eqref{ddeltaV} into~\eqref{rint} leads to
\begin{equation}
\int_{\lambda_0}^\lambda \delta U \delta V d\lambda =
\int_{\delta V(\lambda)}^{\delta V(\lambda_0)} \delta U \delta V
\left( -\frac{d\delta V}{d\lambda} \right)^{-1} d(\delta V)
\leq \int_{\delta V(\lambda)}^{\delta V(\lambda_0)}
\frac{1}{(1-\epsilon)} \frac{1}{\sigma^\prime}\,  d(\delta V) \:.
\end{equation}
This integral remains finite in the limit $\lambda \rightarrow
\infty$ ($\delta V\rightarrow 0$), since $\sigma^\prime \geq
\mathrm{const} > 0$ for small $\eta$ (and thus for small $\delta
V$); hence $r\rightarrow R <\infty$ in the limit $\lambda\rightarrow
\infty$. Finiteness of the mass follows from the finiteness of the
radius.
\end{proof}

In combination with Lemma~\ref{finite}, the results of
Section~\ref{dynanalys} imply theorems that formulate criteria
guaranteeing finiteness of solutions of the Einstein-Vlasov
equations.

Let us first recapitulate the basic assumptions for the theorems: we
consider a distribution function of the type $f = \phi(E) L^{2 l}$
with $l > -1$. Assume that there exists $E_0 > 0$ such that $\phi(E)
=0$ for $E> E_0$, $\phi \in \mathcal{C}^0(0,E_0)$, and $\phi(E) \leq
\mathrm{const} \: (E_0 -E)^{-1 + \epsilon}$ for some $\epsilon >0$
in a neighborhood of $E=E_0$. If $-1<l<-1/2$ we assume, in addition,
that $\phi$ is H\"older continuous with index $\alpha> -l - 1/2$.

\begin{theorem}\label{3theorem}
Suppose there exists $\hat{\eta} > 0$, arbitrarily small, such that $\sigma(\eta)$ satisfies
\begin{equation}
\sigma^\prime \geq \frac{1}{4 + 2 l} + \frac{7 + 4 l}{4 + 2 l} \:\sigma
\end{equation}
for all $\eta \leq \hat{\eta}$.
Then every solution of the static Einstein-Vlasov equations has finite radius
and mass.
\end{theorem}

\begin{remark}
In the asymptotically polytropic case,
provided that $\sigma^\prime|_{\eta(\omega)} \in \mathcal{C}^1[0,1)$,
the condition $\sigma^\prime_0 \geq 1/(4+2 l)$ ($\Leftrightarrow n\leq 3 +l$)
is already sufficient.
\end{remark}

\begin{remark}
Theorem~\ref{3theorem}
is a natural generalization
of~\cite[Theorem 3.1]{reiren00}, where
it is assumed, in our notation, that $l>-1/2$, $n+l > 1$, and $n+l <3$.
\end{remark}

\begin{proof}
The theorem follows immediately from Proposition~\ref{n3prop} and Lemma~\ref{finite}.
\end{proof}

\begin{theorem}\label{5theorem}
Suppose that there exists $\check{\eta}>0$, such that $\sigma(\eta)$ satisfies
\begin{equation}\label{n5coag}
\sigma^\prime \geq \frac{1}{2(3 + 2l)} + k \:\sigma
\end{equation}
for all $\eta \leq \check{\eta}$, where $k = 3 + 2 l + \frac{1}{2}$
if $l \geq -\frac{3}{4}$, and $k =2$ if $ -1 < l < -\frac{3}{4}$.
Then every regular solution of the static Einstein-Vlasov equation
whose central value $\eta_{\mathrm{c}}$ of the potential is smaller
than $\check{\eta}$, i.e., $\eta_c < \check{\eta}$, has finite
radius and mass.
\end{theorem}

\begin{remark}
Finiteness of the radius and the mass also holds for solutions whose
$\alpha$-limit lies on $L_1$, which correspond to solutions that
possess a negative mass singularity.
\end{remark}

\begin{proof}
The theorem follows by combining Proposition~\ref{n5prop} and Lemma~\ref{finite}.
\end{proof}

\begin{remark}
In the asymptotically polytropic case, the condition $n < 5 + 3l$
($\Leftrightarrow \sigma^\prime_0 > 1/(6 + 4l)$) is sufficient
for~\eqref{n5coag} to hold for some $\check{\eta}$,
see~\eqref{n5condiaga}. Therefore, if $n< 5 +3l$, then all regular
solutions with sufficiently small $\eta_{\mathrm{c}}$ have finite
radii.
\end{remark}


\section{Concluding remarks}
\label{concl}

In this paper we have recast the stationary spherically symmetric
Einstein-Vlasov equations for
distribution functions of the form $f=\phi(E)L^{2l}$ into a
three-dimensional dynamical system with compact closure. This
has allowed us to derive new theorems formulating conditions under which 
solutions have finite radii and masses; our theorems naturally extend
previous work, see e.g.,~\cite{reiren00,rensch91} and the Newtonian
analysis in~\cite{heietal06}.

In the present paper we have focused on
distributions that can be bounded by asymptotic polytropes in the
low $\eta$ regime. Additional results can be obtained if one
also makes restrictions that lead to controlled behavior in the
large $\eta$ regime; e.g., one can investigate classes of distributions 
that yield asymptotically linear relations between the pressures and $\rho$,
which in turn may make it possible to obtain theorems about
mass-radius diagrams of the type given in~\cite{heietal03} for
perfect fluids with asymptotically linear equations of state in the
high pressure regime.

The theorems presented in this article 
may be less sharp than one would wish, since the natural kinetic 
assumptions
imply non-trivial restrictions on the effective implicit equation
of state that have not been investigated here.
The connection between the distribution function and
the effective equation of state is given via the integrals
of Section~\ref{introduction}; it is an interesting question whether
it is possible
obtain more comprehensive information about this relation, e.g., 
by finding classes of functions $\phi(E)$ that lead
to certain types of equations of state.
Simple results in this direction are due to Fackerell~\cite{fac68}: 
it could be excluded that models with constant energy density possess
a kinetic interpretation. In
this context one should note that the assumption $l\neq 0$ implies that 
the density and the pressures depend on both $r$ and $\eta$; 
however, the function $\sigma$ appearing on the r.h.s.\ of
the equations is a function of $\eta$ alone.
Hence the mathematical correspondence 
between fluids with equations of state and kinetic
models amounts to a correspondence in $\sigma(\eta)$ alone --- which
equations of state yield $\sigma(\eta)$ with kinetic
interpretations?

\begin{appendix}

\section{Differentiability and asymptotic behavior}
\label{appdiff}

The r.h.\ sides of the dynamical systems~\eqref{uvomeq}
and~\eqref{dynsys} contain the function $\sigma(\eta)$ and its
derivative. To analyze the differentiability properties of these
function we investigate the key quantities
\begin{equation}\label{gmagain}
g_{i}(\eta)= \int_{e^{-\eta }}^{1}\, \phi(\cE )\,
\Big[\textfrac{1}{2}(\cE ^{2}-e^{-2\eta})\Big]^{i}\,d\cE\:,
\end{equation}
on which $\sigma(\eta)$ is built, see~\eqref{gidef} and~\eqref{sig}.

\begin{lemma}\label{deriv}
Let $\phi: (0,1) \ni \cE \mapsto \mathbb{R}$ be 
non-negative.
Assume that $\phi \in \mathcal{C}^{k}(0,1)$ and that $\big|(1-\cE)^k
\phi^{(k)}(\cE)\big| \leq \mathrm{const}\, (1-\cE)^{-1+\epsilon}$
for some $\epsilon>0$ in a neighborhood of $\cE=1$. Then, for $i\geq
0$, $g_i(\eta) \in \mathcal{C}^{k+1}(0,\infty)$ and
\begin{equation}\label{dgdeta}
\frac{d}{d\eta} g_i(\eta) = i e^{-2 \eta} g_{i-1}
\quad(i>0)\:,\qquad \frac{d}{d\eta} g_0(\eta) = e^{-\eta}
\phi(e^{-\eta})\:.
\end{equation}
In the case $-1<i<0$, $g_i(\eta) \in \mathcal{C}^{k}(0,\infty)$;
however, if the additional H\"older continuity condition $(1-\cE)^k
\phi^{(k)}(\cE) \in \mathcal{C}^\alpha(0,1)$ with $\alpha > -i$
holds, then again $g_i(\eta) \in \mathcal{C}^{k+1}(0,\infty)$.
\end{lemma}

\begin{proof}
The idea of the proof is to transform the function $g_i$ to a form
that corresponds to~Equation~(A1) of~\cite{heietal06} and to use the
results of~\cite[Appendix A]{heietal06}. Employing an adapted
variable $\omega$ instead of $\eta$, cf.~\eqref{omegaeta}, the
function $g_i$ can be regarded as a function $g_i(\omega)$; we set
\begin{equation}
\omega = 1 - e^{-\eta} \:;
\end{equation}
proving differentiability in $\eta$ amounts to proving
differentiability in $\omega$. Note that $0< \omega <1$ since
$0<\eta<\infty$. Replacing $\cE$ by $x = \omega^{-1} (1-\cE)$ in the
integral then transforms~\eqref{gmagain} to
\begin{equation}\label{gmcomp}
g_i(\omega) = \omega^{i+1} \int_0^1 \phi(1-\omega
x)\,G^i(\omega,\omega x) \, (1-x)^{i} \,dx\:,
\end{equation}
where $G(\nu,\tau) := 1 - \frac{\nu}{2}  -\frac{\tau}{2}$ is a
smooth, positive, and bounded function on $(\nu,\tau) \in [0,1)
\times [0,1)$; in particular, $x\mapsto G^i(\omega,\omega x)$ is
smooth, positive, and bounded on $x\in[0,1]$ for all $\omega$.
Define
\begin{equation}\label{gnhat}
\hat{g}_i(\nu,\omega) := \omega^{i+1} \int_0^1 \phi(1-\omega
x)\,G^i(\nu,\omega x)\, (1-x)^{i} \,dx\:;
\end{equation}
evidently, $g_i(\omega) = \hat{g}_i(\omega,\omega)$. Introduce
$\hat{\phi}_i$ according to
$\hat{\phi}_i(\nu, \tau) = \phi(1 - \tau) G^i(\nu,\tau)$\:.
By assumption, $\tau \mapsto \hat{\phi}_i(\nu,\tau)$ is measurable,
non-negative, bounded on compact subsets of $(0,1)$, and
$\mathcal{C}^k$; furthermore, for small $\tau$, $\big|\tau^k
\partial_\tau^{\,k} \hat{\phi}_i(\nu,\tau)\big| \leq
\mathrm{const}\: \tau^{-1+\epsilon}$ for some $\epsilon>0$. By using
$\hat{\phi}_i$, Eq.~\eqref{gnhat} becomes
\begin{equation}
\hat{g}_i(\nu,\omega) = \omega^{i+1} \int_0^1
\hat{\phi}_i(\nu,\omega x)\, (1-x)^{i} \,dx \:.
\end{equation}
%
Therefore, $\omega \mapsto \hat{g}_i(\nu,\omega)$ is of the same
form as Equation~(A1) of~\cite{heietal06}. Using the results
of~\cite[Appendix A]{heietal06} we conclude that $\omega\mapsto
\hat{g}_i(\nu,\omega)$ is $\mathcal{C}^{k+1}$ (where the additional
requirement that the function $(1-\cE)^k \phi^{(k)}$ be H\"older
continuous enters in the case $-1<i<0$). Since $\nu\mapsto
\hat{g}_i(\nu,\omega)$ is smooth we find that $\omega \mapsto
\hat{g}_i(\omega,\omega) = g_i(\omega)$ is $\mathcal{C}^{k+1}$, as
claimed. A straightforward computation based on the relation
\begin{equation}
\frac{d}{d\omega} g_i(\omega) = (\partial_\nu
\hat{g}_i)(\omega,\omega) + (\partial_\omega
\hat{g}_i)(\omega,\omega)
\end{equation}
yields Eq.~\eqref{dgdeta}.
\end{proof}

\begin{remark}
In the case $-1<i<0$ the derivative of $g_i$ is given by
\begin{equation}
\frac{dg_i }{d\eta }(\eta )= \omega^i 
e^{-(i+1)\eta }\phi (e^{-\eta})
+i e^{-2\eta } \omega^i \, \int_{0}^{1}
\Psi (\eta ,x) (1-x)^{\delta -1}dx, 
\end{equation}
where $\delta = \alpha + i >0$ and $\omega = 1 -e^{-\eta}$;
since $\phi$ is H\"older continuous with index $\alpha$, the
function 
$\Psi (\eta ,x)=
\left[\phi\big(e^{-\eta} + \omega (1-x)\big)
\big(e^ {-\eta}+\frac{\omega}{2} (1-x)\big)^{i-1}-
\phi (e^{-\eta}) (e^{-\eta })^{i-1}\right] (1-x)^{-\alpha}$
is bounded and the integral is well-defined.
The derivation of this formula is analogous to the considerations
of~\cite[Appendix A]{heietal06}; we omit the details here.
\end{remark}

Lemma~\ref{deriv} implies the following corollary:

\begin{corollary}\label{Dcol}
Let $\phi: (0,1) \ni \cE \mapsto \mathbb{R}$ be 
non-negative.
Assume that $\phi \in \mathcal{C}^{k}(0,1)$ and that $\big|(1-\cE)^k
\phi^{(k)}(\cE)\big| \leq \mathrm{const}\, (1-\cE)^{-1+\epsilon}$
for some $\epsilon>0$ in a neighborhood of $\cE=1$. Then
\begin{equation}
g_i(\eta) \in \mathcal{C}^{[i]+k+1}(0,\infty)\:,
\end{equation}
where $[i]$ denotes the largest integer less or equal to $i$. For
$i\not\in \mathbb{N}$, if in addition $(1-\cE)^k \phi^{(k)}(\cE) \in
\mathcal{C}^\alpha(0,1)$ with $\alpha > 1+[i]-i$, then $g_i(\eta)
\in \mathcal{C}^{[i]+k+2}(0,\infty)$.
\end{corollary}

\begin{proof}
Suppose that $\mathbb{N}\not\ni i >0$. Applying~\eqref{dgdeta}
iteratively $[i]$ times we find that $d^{[i]} g_i/d \eta^{[i]}
\propto g_{i-[i]}$. Since $i-[i]>0$, $g_{i-[i]}$ is in
$\mathcal{C}^{k+1}$ by Lemma~\ref{deriv}; accordingly, $g_{i} \in
\mathcal{C}^{[i]+k+1}$. The case $i\in\mathbb{N}$ is analogous:
$i=[i]$, $d^{i} g_i/d \eta^{i} \propto g_0$, and $g_0 \in
\mathcal{C}^{k+1}$. If in addition $(1-\cE)^k \phi^{(k)}(\cE) \in
\mathcal{C}^\alpha(0,1)$ with $\alpha > 1+[i]-i$
($i\not\in\mathbb{N}$), then
 $d^{[i]+1} g_i/d \eta^{[i]+1} \propto g_{i-[i]-1}$, which is
$\mathcal{C}^{k+1}$ by assumption; from this the claim follows.
\end{proof}

Recall from~\eqref{sigmaform} that
\begin{equation}\label{sigeta}
\sigma(\eta) = \left[3+2l + \frac{d}{d\eta} \log
g_{l+3/2}(\eta)\right]^{-1}\:.
\end{equation}
Accordingly,
\begin{equation}\label{dsigeta}
\frac{d\sigma}{d\eta} = - \sigma^2 \frac{d^2}{d\eta^2} \log
g_{l+3/2}(\eta)\:.
\end{equation}
Under the assumptions of Corollary~\ref{Dcol} we see that
$g_{l+3/2}$ is in $\mathcal{C}^{[l+3/2] + k + 1}$ so that
$d\sigma/d\eta$ is in $\mathcal{C}^{[l+3/2] + k - 1}$ (and in
$\mathcal{C}^{[l+3/2] +k}$ when the H\"older condition is
satisfied). In particular, if $l\geq -1/2$, then the assumptions
$\phi \in \mathcal{C}^{0}(0,1)$, $|\phi(\cE)| \leq \mathrm{const}\,
(1-\cE)^{-1+\epsilon}$ for some $\epsilon>0$ as $\cE\rightarrow 1$,
i.e., $k=0$ in Corollary~\ref{Dcol}, suffice to obtain
$g_{l+3/2} \in \mathcal{C}^2$ and $d\sigma/d\eta \in \mathcal{C}^0$;
in the case $l<-1/2$, the
additional H\"older condition must be imposed. Higher
differentiability of $\phi$ (larger $k$) leads to higher
differentiability of $d\sigma/d\eta$.

\begin{definition}
A function $g_i(\eta)$, $i>-1$, is said to possess
\textit{asymptotically polytropic} behavior if there exists
$\mathbb{R} \ni n>1/2$ such that
\begin{equation}\label{gmbehav}
g_i(\eta) = \mathrm{const} \:  \eta^{i+n-1/2} \left( 1 + o(1)
\right)
\end{equation}
as $\eta\rightarrow 0$. We require the derivatives of $g_i$, if they
exist, to be of the same form with a lowered exponent, i.e., $d^k
g_i/d \eta^k \propto \eta^{i+n-1/2-k} \left( 1 + o(1) \right)$.
\end{definition}

Functions $g_i$ with asymptotically polytropic behavior are
generated by asymptotically polytropic distribution functions:

\begin{definition}
A distribution $\phi(\cE)$ (which is at least $\mathcal{C}^0$)
is said to possess asymptotically polytropic behavior as $\cE\rightarrow 1$,
if there exists
$\mathbb{R}\ni n>1/2$ such that
\begin{equation}\label{asypolphi2}
\phi(\cE) = \phi_- (1-\cE)^{n-3/2} \, \left(1 + o(1) \right)
\end{equation}
as $\cE\rightarrow 1$; here, $\phi_- >0$ is a constant.
\end{definition}

\begin{lemma}\label{asyasy}
Distributions with asymptotically polytropic behavior generate functions
$g_i(\eta)$ with asymptotically polytropic behavior.
\end{lemma}

\begin{proof}
By Corollary~\ref{Dcol}, $g_i \in \mathcal{C}^{[i]+1}$, since
$\phi$ is continuous.
To prove the claim of the lemma we have to
show that $g_i$ (and analogously its
derivatives) behave according to~\eqref{gmbehav}. To that end set
$\phi(\cE) = \phi_1(\cE) + \phi_2(\cE)$ with
\begin{equation}
\phi_1(\cE) = \phi_- (1-\cE)^{n-3/2} \:, \qquad \phi_2(\cE) =
\varphi(\cE)  (1-\cE)^{n-3/2}\:,
\end{equation}
where $\varphi(\cE) = o(1)$ as $\cE\rightarrow 1$. By definition,
\begin{equation}
g_i(\eta) = \int_{e^{-\eta}}^1 \left(\phi_1(\cE) +
\phi_2(\cE)\right)\Big[\textfrac{1}{2}(\cE
^{2}-e^{-2\eta})\Big]^{i}\,d\cE = \int_{e^{-\eta}}^1
\left(\phi_1(\cE) + \phi_2(\cE)\right)\, K^{i}\,d\cE\:,
\end{equation}
where we have set $K = (\cE ^{2}-e^{-2\eta})/2$. The first integral
can be computed explicitly by using hypergeometric functions; series
expansion of the result leads to
\begin{equation}
\int_{e^{-\eta}}^1 \phi_1(\cE) K^{i}d\cE = \phi_- \int_{e^{-\eta}}^1
(1-\cE)^{n-3/2} K^i d\cE = \mathrm{const}\: \eta^{i+n-1/2} \left(1
+o(1) \right)
\end{equation}
as $\eta\rightarrow 0$. For $\eta$ small enough, the second integral
can be estimated as
\begin{align}\nonumber
\Big|\int_{e^{-\eta}}^1 \phi_2(\cE) K^{i}\,d\cE \Big|
& = \Big|\int_{e^{-\eta}}^1 \varphi(\cE) (1-\cE)^{n-3/2} K^i d\cE \Big| \\[0.5ex]\nonumber
& \leq \mathrm{const}\: \Big(\max \{ \varphi(\cE) \,|\,
\cE\in[e^{-\eta},1] \} \Big) \:\,  \eta^{i+n-1/2} =
o(\eta^{i+n-1/2})\:.
\end{align}
Combining the results shows that
\begin{equation}
g_i(\eta) = \mathrm{const}\: \eta^{i+n-1/2} \left( 1 + o(1) \right)
\:.
\end{equation}
By Corollary~\ref{Dcol}, $g_i \in \mathcal{C}^{[i]+1}$. To prove the
lemma it remains to investigate the derivatives of $g_i$. If $i>0$,
the derivative of $g_i(\eta)$ is given by~\eqref{dgdeta}, i.e.,
\begin{equation}
\frac{d g_i}{d\eta} =i\,e^{-2\eta}\,g_{i-1} = \mathrm{const}\:
\eta^{i+n-3/2} \left( 1 + o(1) \right)\:,
\end{equation}
which shows that the derivative is asymptotically polytropic (with a
lower exponent, as required). By iteration we obtain that all
derivative behave correctly. If $i=0$, we have
\begin{equation}
\frac{d g_0}{d\eta} = e^{\eta} \phi(e^{-\eta}) = e^{-\eta}
\eta^{n-3/2} \left( 1 + o(1) \right) = \eta^{n-3/2} \left( 1 + o(1)
\right)\,
\end{equation}
which is again the required behavior. If $i<0$, $g_i$ is
differentiable only if $\phi$ satisfies the additional H\"older
continuity condition of~Corollary~\ref{Dcol}. Also in this case it
can be shown that the derivative behaves in an asymptotically
polytropic manner; we omit the details here.
\end{proof}

\begin{remark}
If $\phi(\cE)\in\mathcal{C}^k$ ($k>0$) and $\big|(1-\cE)^k
\phi^{(k)}(\cE)\big| \leq \mathrm{const}\, (1-\cE)^{-1+\epsilon}$,
then higher derivatives of $g_i(\eta)$ exist,
cf.~Corollary~\ref{Dcol}. In analogy to Lemma~\ref{asyasy} it can be
proved that these higher derivatives exhibit the correct
asymptotically polytropic behavior provided that $(1-\cE)^k
\phi^{(k)}(\cE)$ is asymptotically polytropic like $\phi(\cE)$
in~\eqref{asypolphi2}.
\end{remark}

\begin{lemma}[The asymptotically polytropic case]\label{sigsiglemma}
Let $\phi(\cE) \in \mathcal{C}^0(0,1)$ be an asymptotically polytropic
distribution function, see~\eqref{asypolphi2}.
(In the case $l<-1/2$, we require in addition that
$\phi(\cE)$  be H\"older continuous
with index $\alpha > -l -1/2$.)
Then $g_{l+3/2}(\eta)$ is $\mathcal{C}^2$ and asymptotically
polytropic as $\eta\rightarrow 0$, and
\begin{subequations}\label{sigsig}
\begin{align}
&\lim_{\eta\rightarrow 0} \sigma(\eta) = 0 \\
&\lim_{\eta\rightarrow 0} \frac{d\sigma}{d\eta} \:= \frac{1}{1+ l
+n}\:.
\end{align}
\end{subequations}
\end{lemma}

\begin{proof}
Since $\phi \in \mathcal{C}^0$, Corollary~\ref{Dcol} implies
that $g_{l+3/2}(\eta)$ is $\mathcal{C}^2$ (where the H\"older continuity
condition is needed in the case $l<-1/2$).
Asymptotically polytropic behavior of $g_{l+3/2}$ (and its two
derivatives) follows from
Lemma~\ref{asyasy}.
Accordingly, we have
\begin{equation}
\log g_{l+3/2} = \mathrm{const} + (l+n+1) \log \eta + \log (1 +o(1))
\:,
\end{equation}
where the derivative of $o(1)$ is of the order $o(\eta^{-1})$ as
$\eta\rightarrow 0$; hence, from~\eqref{sigeta},
\begin{equation}
\sigma = \left[ 2 l + 3 + (l+n+1) \eta^{-1} \big(1 +
o(1)\big)\right]^{-1} \rightarrow 0 \quad (\eta\rightarrow 0)\:.
\end{equation}
Furthermore, Eq.~\eqref{dsigeta} results in
\begin{equation}
\frac{d\sigma}{d\eta} = - \sigma^2 \left[ -(l+n+1) \eta^{-2}
\big(1+o(1)\big) \right] \rightarrow \frac{1}{l+n+1} \quad
(\eta\rightarrow 0),
\end{equation}
which proves the claim.
\end{proof}

\section{Global existence and uniqueness of solutions}
\label{appexist}

If $\sigma^\prime$ (as a function of $\Omega$) is in
$\mathcal{C}^1(0,1)$, the r.h.s.\ of the dynamical
system~\eqref{dynsys} is $\mathcal{C}^1$; existence and uniqueness
of solutions follows directly from the theory of dynamical systems.
These issues are less trivial when only continuity is imposed on
$\sigma^\prime$. In the following we establish (global) existence
and uniqueness of solutions of the dynamical system~\eqref{dynsys}
on the state space ${\mathbf X}=(U,V,\Omega) \in [0,1]^2\times
(0,1)$ under the assumption that $\sigma^\prime$ be continuous on

$(0,1)$.

We begin by introducing a new independent variable $\zeta$
according to
\begin{equation}
\frac{d\lambda}{d\zeta}=H^{-1}=[(1-U+\sigma U)V]^{-1}\:.
\end{equation}
Then equation~(\ref{dOmega}) becomes
$d\Omega/d\zeta = -\Omega (1-\Omega ) F(\Omega)$, where $F(\Omega)$
is a positive function bounded away from zero for all $\Omega \in
[0,1]$. Therefore, this equation has a unique global (in $\zeta$)
solution for any initial value $\Omega(0) \in (0,1)$. In particular,
$\Omega(\zeta)$ is a monotonic function mapping the interval
$(-\infty,\infty)$ to $(0,1)$. Eqs.~(\ref{dU}) and~(\ref{dV}) take
the form
\begin{subequations}
\begin{align}
\label{dUmu}
\frac{dU}{d\zeta} & = \frac{U(1-U)}{V(1-U+\sigma
U)}\left[[3 + 2 l - 2(2+ l)U](1-V) -
\left(1+\sigma - \frac{d\sigma}{d\eta}\right)\,H\right]\:,\\
\label{dVmu}
\frac{dV}{d\zeta} & =  \frac{1-V}{1-U+\sigma
U}\left[(2U-1)(1-V+2\sigma V) + \frac{d\sigma}{d\eta}\,H\right]\:,
\end{align}
\end{subequations}
where $\sigma$ and $\sigma^\prime$ are now regarded as functions of
$\zeta$. For any initial data $(U(0),V(0))\in [0,1]\times (0,1]$ the
above system has a unique (local in $\zeta$) solution. This follows
immediately from the Picard-Lindel\"of theorem, since the r.h.s.\ is
continuous in $\zeta$ and smooth in $U$ and $V$. Translating back to
the original variables, we have achieved existence and uniqueness of
solutions in the space $[0,1]\times (0,1]\times (0,1)$; a priori,
existence of solution is local in $\lambda$.

Solutions with initial data in the space $[0,1]\times (0,1]\times
(0,1)$ cannot reach the side face $V=0$ in finite $\lambda$-time.
This is a direct consequence of the fact that $V^{-1} d V/d\lambda$
is bounded away from zero as $V\rightarrow 0$ (where we use that
$U\not\rightarrow 1/2$). On the side face $V=0$ itself, existence
and uniqueness of solutions is obvious, since the induced system has
a simple structure. Combining these facts we have established (local
in time $\lambda$) existence and uniqueness of solutions on the
state space $[0,1]^2 \times (0,1)$.

To obtain global existence of solutions we proceed as follows.
Suppose that a (local) solution has been extended to a maximal
interval $(\lambda_-,\lambda_+)$ of existence. Assume that
$\lambda_+ \neq \infty$. Since the solution ceases to exist for
$\lambda \geq \lambda_+$, it must leave the state space as $\lambda
\rightarrow \lambda_+$, i.e., $\Omega(\lambda) \rightarrow 0$ as
$\lambda \rightarrow \lambda_+$ (note that $\Omega$ must be
decreasing). However, $\Omega(\lambda)$ satisfies the differential
inequality $d\Omega/d\lambda \geq - F \Omega \geq -\mathrm{const}\,
\Omega$, therefore $\Omega(\lambda) \rightarrow \mathrm{const} > 0$
as $\lambda\rightarrow \lambda_+$, which is a contradiction. Hence,
$\lambda_+ = \infty$. Since we can show in an analogous way that
$\lambda_- = -\infty$, the solution exists globally for all
$\lambda$.

\section{Center manifold analysis}
\label{appcenter}

Assume that $\sigma^\prime$ (as a function of $\Omega$) is in
$\mathcal{C}^1[0,1)$ so that we have an asymptotically polytropic
$C^1$-differentiable dynamical system on the state space $[0,1]^2
\times [0,1)$. Then the following Lemma holds:

\begin{lemma}\label{centerlemma}
The fixed point $B_1=(0,\frac{1}{1+\sigma^\prime_0},0)$
is an attractor for interior orbits if and only if $\sigma^\prime_0
<\frac{1}{4+2l}$.
\end{lemma}

\begin{proof}
If $\sigma_0^\prime < 1/(4+2l)$, the fixed point $B_1$
is hyperbolic: two eigenvalues are negative, one is positive.
Analyzing the directions of the associated eigenvectors we find
that there exists a one-parameter family of orbits that
converges to $B_1$ as $\lambda\rightarrow \infty$.

Now assume that $\sigma^\prime_0 >\frac{1}{4+2l}$. In this case the
fixed point $B_1$ is a hyperbolic saddle (see Table~\ref{fixtab})
with the $\Omega=0$ plane as unstable manifold; it immediately
follows that $B_1 $ attracts no interior orbits.

The case $\sigma^\prime_0=\frac{1}{4+2l}$ is more delicate to
analyze, since one of the eigenvalues is zero in this case; we make
use of center manifold theory in our analysis. We define new adapted
variables according to
\begin{equation}
x:=(4+2l)u\:, \qquad y:=(4+2l)u+v-(4+2l)\:,\qquad z:=\eta\:;
\end{equation}
using these variables transforms the dynamical system~\eqref{uvomeq} to a system of the form
\begin{equation}
\frac{dx}{d\xi }=N_{x}(x,y,z)\:, \quad \frac{dy}{d\xi}=
y+N_{y}(x,y,z)\:, \quad \frac{dz}{d\xi }=-z+N_{z}(x,y,z)\:,
\end{equation}
where
\begin{equation*}
N_{x}(x,y,z)=x \left[3+2l-\sigma^\prime_0 x - 
\big(1+\sigma(z)-\sigma^\prime(z)\big)\big(1+\sigma^\prime_0 \sigma(z) x\big) 
\left(\frac{1}{\sigma^\prime_0}+y-x\right)\right].
\end{equation*}
%
The center manifold can be represented by the local graph
$(x,h_{y}(x),h_{z}(x))$, which satisfies the system of equations
\begin{subequations}\label{sigsig2}
\begin{align}
& \partial _{x}h_{y}(x)\frac{dx}{d\xi }(x,h_{y}(x),h_{z}(x))=
\frac{dy}{d\xi }(x,h_{y}(x),h_{z}(x))\:,\\
& \partial _{x}h_{z}(x)\frac{dx}{d\xi
}(x,h_{y}(x),h_{z}(x))=\frac{dz}{d\xi }(x,h_{y}(x),h_{z}(x))\:,
\end{align}
\end{subequations}
and the tangency conditions $h_{i}(0)=0$, $\partial_{x}h_{i}(0)=0$
($i=y,z$). By using expansions of the functions we are able to solve
the system up to any desired order $x^{n}$; we obtain
$h_{z}(x)\equiv 0$ and $h_{y}(x)=-\frac{2+2l}{4+2l}x^2
-\frac{(8+6l)(2+2l)}{(4+2l)^2}x^3 +O(x^4 )$. The center manifold
reduction theorem states that the full nonlinear system is locally
equivalent to the flow of the reduced system given by $dx/d\xi =
N_x(x,h_y(x),h_z(x))$, $d y/d\xi =y$, $d z/d\xi =-z$, i.e.,
\begin{subequations}\label{sigsig3}
\begin{align}
& \frac{dx}{d\xi }= \frac{1+l}{2+l}x^2 -\frac{(2+2l)(3+2l)}{(4+2l)^2
}x^3 + O(x^4 )\:,\\
& \frac{dy}{d\xi }=y\:, \\
& \frac{dz}{d\xi }=-z\:.
\end{align}
\end{subequations}
The interior orbits are given by $x>0$, $y\in\mathbb{R}$ and $z>0$;
clearly, the flow of the reduced system prevents such orbits from
converging to $B_{1}$ so that $B_1$ does not attract interior
orbits. This concludes the proof of the lemma.
\end{proof}

The proof of Proposition~\ref{nequals3prop} is now straightforward:

\textbf{Proposition~\ref{nequals3prop}.}
\textit{Assume that there exists a variable $\omega$ such that $\sigma^\prime|_{\eta(\omega)}$
is $\mathcal{C}^1[0,\epsilon)$ (for some $\epsilon >0$).
If
\begin{equation}
\sigma_0^\prime \geq \frac{1}{4 + 2l}\:,\qquad \text{or, equivalently,}\qquad
n \leq 3 + l\:,
\end{equation}
then the $\omega$-limit of every interior orbit lies
on $(\Omega = 0) \cap (V=1)$.}

\begin{proof}
The case $\sigma^\prime_0 > 1/(4+2l)$ ($\Leftrightarrow n < 3 +l$)
follows readily from Proposition~\ref{n3prop}. If $\sigma^\prime_0 =
1/(4+2l)$ ($\Leftrightarrow n = 3 +l$), there are a priori two
attractors (which follows from analyzing the flow on $\Omega =0$
with the aid of the monotonic function $Z$, cf.~\eqref{Zfctdef}): a
fixed point on $(\Omega = 0) \cap (V=1)$ or the fixed point $B_1$.
By Lemma~\ref{centerlemma} the latter possibility is excluded, which
proves the proposition.
\end{proof}

\section{The monotonicity principle}
\label{appmon}

In this section we prove an extension of the monotonicity
principle presented in~\cite{LeBlanc/Kerr/Wainwright:1995}) (see also~\cite{waiell97}).

Consider an open set $\mathcal{S}\subset \mathbb{R}^n$ and its closure $\bar{\mathcal{S}}$.
Let $\bar{s}$ be a point in $\bar{\mathcal{S}}$.
We denote by $B_\epsilon(\bar{s})$ an open ball with radius $\epsilon$ and center $\bar{s}$,
and by $\mathcal{B}_\epsilon(\bar{s})$ its intersection with $\mathcal{S}$, i.e.,
$\mathcal{B}_\epsilon(\bar{s}) = B_\epsilon(\bar{s}) \cap \mathcal{S}$.

Let $Y: \mathcal{S} \rightarrow (0,1)$ denote a continuous function.
For every $\bar{s} \in \bar{\mathcal{S}}$ we define
\begin{equation}
Y_-(\bar{s}) := \lim_{\epsilon\rightarrow 0} \:\inf_{x \in \mathcal{B}_\epsilon(\bar{s})} Y(x)
\:,\qquad
Y_+(\bar{s}) := \lim_{\epsilon\rightarrow 0} \:\sup_{x \in \mathcal{B}_\epsilon(\bar{s})} Y(x)\:;
\end{equation}
the resulting interval $[Y_-(\bar{s}), Y_+(\bar{s})]$ we call the ``limit set'' of $Y(x)$ as $x\rightarrow \bar{s}$:
\begin{equation}
\primelim_{x\rightarrow \bar{s}} Y(x) := [Y_-(\bar{s}), Y_+(\bar{s})]\:.
\end{equation}
Note that $Y_-$ ($Y_+$) can be any number from $0$ to $1$
(provided that $Y$ is onto).

\begin{remark}
Suppose that the set $\mathcal{S}$ possesses a ``regular shape'', i.e., assume that
for each $\bar{s} \in \bar{\mathcal{S}}$
there exists $\delta > 0$ such that
$\mathcal{B}_\epsilon(\bar{s})$ 
is connected for all $\epsilon < \delta$. (This is the case, e.g., for
convex sets $\mathcal{S}$).
Under this assumption,
for each $a \in \primelim_{x\rightarrow \bar{s}} Y(x)$ there exists
a sequence $(x_n)_{n\in\mathbb{N}}$, $x_n \in \mathcal{S}$ $\forall n$,
such that $a = \lim_{n\rightarrow \infty} Y(x_n)$. This fact justifies the
notation $\primelim_{x\rightarrow \bar{s}} Y(x)$ for the interval $[Y_-(\bar{s}), Y_+(\bar{s})]$.
\end{remark}

\begin{remark}
The set $\primelim_{x\rightarrow \bar{s}} Y(x)$
depends in a continuous way on $\bar{s}$:
if  
$\big(\primelim_{x\rightarrow \bar{s}} Y(x)\big) \cap [a-\delta, a +\delta] = \emptyset$
for some $a \in \mathbb{R}$ and $\delta > 0$, then
for every $\delta^\prime < \delta$ there exists a neighborhood $U(\bar{s})$ of $\bar{s}$ such
that $\big(\primelim_{x\rightarrow \bar{\sigma}} Y(x)\big) \cap [a-\delta^\prime, a +\delta^\prime] = \emptyset$
for all $\bar{\sigma} \in U(\bar{s})$.
\end{remark}

\begin{remark}
For a function $\tilde{Y}$ that possesses the range $(-\infty, \infty)$
the limit set $\primelim \tilde{Y}$ is defined in complete analogy;
in this case, $\primelim \tilde{Y}$ is a subset of $[-\infty, \infty]$.
Note that $\tilde{Y}$ can be manipulated by using a
diffeomorphism between $(0,1)$ and $(-\infty, \infty)$
to obtain the original set-up.
\end{remark}
 
Consider a dynamical system
\begin{equation}\label{dysy}
\frac{d x}{d t} = f(x) 
\end{equation}
that is defined on a future invariant open set $\mathcal{S} \subseteq \mathbb{R}^n$;
in the case when $f$ is merely continuous, we require existence and uniqueness of solutions 
of~\eqref{dysy} to hold.
We say that $f(x)$ does not diverge as $x$ approaches the boundary $\partial \mathcal{S}$
of $\mathcal{S}$, if the condition $\pm \infty \not \in \primelim_{x\rightarrow \bar{s}} f(x)$
holds for all $\bar{s} \in \partial\mathcal{S}$.
(In fact, in the following theorem we will meet the slightly less restrictive assumption that
$\pm \infty \not \in \primelim_{x\rightarrow \bar{s}} f(x)$,
$\forall \bar{s} \in \partial\mathcal{S}\backslash (\partial\mathcal{S})_{\mathrm{inf}}$.)

Suppose that there exists a $\mathcal{C}^1$ function $Z:\mathcal{S} \rightarrow \mathbb{R}$ 
that is strictly monotonically increasing along all orbits in $\mathcal{S}$.
We define the subset $(\partial \mathcal{S})_{\mathrm{inf}}$ of $\partial \mathcal{S}$ as
\begin{equation}
(\partial \mathcal{S})_{\mathrm{inf}} =
\left \{ \bar{s} \in \partial \mathcal{S} \:|\: \primelim_{x \rightarrow \bar{s}} Z(x) = \{\inf_{\mathcal{S}} Z\} \right\}\:.
\end{equation}
Based on the definitions and concepts introduced above we are now able to
state and prove the following theorem:

\begin{theorem}[Monotonicty principle]\label{monprinc}
Consider a dynamical system $d x/d t = f(x)$ on the future-invariant
open set $\mathcal{S}\subseteq \mathbb{R}^n$;
$f(x)$ is required to be non-divergent on $\partial\mathcal{S}\backslash (\partial\mathcal{S})_{\mathrm{inf}}$.
Suppose that there exists a $\mathcal{C}^1$ function $Z: \mathcal{S} \rightarrow \mathbb{R}$ 
that is strictly monotonically increasing along orbits in ${\mathcal S}$,
whereby the derivative $\dot{Z} = \nabla Z \cdot f(x)$ is a function $\dot{Z}: \mathcal{S} \rightarrow (0,\infty)$.
Then for all $x\in{\mathcal S}$,
\begin{equation}
\omega(x) \subseteq \left\{\bar{s} \in \partial\mathcal{S} \backslash (\partial \mathcal{S})_{\mathrm{inf}} \:\big|\:
\big(\primelim\limits_{x\rightarrow \bar{s}} Z(x) \ni \sup\limits_{\mathcal S} Z\big)  \:\vee\:
\big(\primelim\limits_{x\rightarrow \bar{s}} \dot{Z}(x)\ni 0\big) \:\right\}\:,
\end{equation}
\end{theorem}

\begin{proof}
Since $Z$ is strictly increasing along all orbits in $\mathcal{S}$,
the infimum $\inf Z$ (and the supremum $\sup Z$) is not assumed in $\mathcal{S}$.
For each $x\in\mathcal{S}$ we have $Z(x)> \inf Z$,
and hence $\omega(x) \not\in (\partial \mathcal{S})_{\mathrm{inf}}$.

Assume that there exists $x\in{\mathcal S}$ such that
\begin{equation}
\omega(x) \cap  \left\{\bar{s} \in \partial\mathcal{S} \backslash (\partial \mathcal{S})_{\mathrm{inf}} \:\big|\:
\big(\primelim\limits_{x\rightarrow \bar{s}} Z(x) \not\ni \sup\limits_{\mathcal S} Z\big)  \:\wedge\:
\big(\primelim\limits_{x\rightarrow \bar{s}} \dot{Z}(x)\not\ni 0\big) \:\right\}
\neq \emptyset \:,
\end{equation}
and let $\bar{s}$ be an element of this set.
Consequently, there exists a sequence $(t_n)$, $n\in\mathbb{N}$, such that
$t_n \rightarrow \infty$ as $n\rightarrow\infty$ and $\lim_{n\rightarrow\infty} \phi_{t_n}(x) = \bar{s}$
(where $\phi_t$ denotes the flow of the dynamical system).
Since $Z$ is monotonically increasing along orbits, the sequence
$Z(\phi_{t_n}(x))$, $n\in\mathbb{N}$, is strictly monotonically increasing.
The limit of this sequence is finite, $\lim_{n\rightarrow\infty} Z(\phi_{t_n}(x)) < \infty$,
because $\primelim_{x\rightarrow \bar{s}} Z(x) \not\ni \sup_{\mathcal S} Z$
(and $\primelim_{x\rightarrow \bar{s}} Z(x)$ is a closed interval).
In contrast, the limit $\lim_{n\rightarrow\infty} \dot{Z}(\phi_{t_n}(x))$ need not exist;
since $\primelim\limits_{x\rightarrow \bar{s}} \dot{Z}(x)\not\ni 0$,
there exists $\varepsilon >0$ such that
$\dot{Z}(\phi_{t_n}(x)) \geq 2 \varepsilon$ for sufficiently large $n \in \mathbb{N}$.
Similarly, by assumption, there exists $F \in (0,\infty)$ such that
$|f(\phi_{t_n}(x))| \leq \,F/2$ for sufficiently large $n$.

Choose a neighborhood $U(\bar{s})$ of $\bar{s}$ such that
\begin{equation}
\primelim_{x\rightarrow\bar{\sigma}} \dot{Z}(x) \geq \varepsilon \quad\mbox{and}\quad
\primelim_{x\rightarrow\bar{\sigma}} |f(\phi_{t_n}(x))| \leq F
\end{equation}
for all $\bar{\sigma} \in U(\bar{s})$.
Let $d$ be defined as $2 d = \inf_{s\in {\mathcal S}\backslash U(\bar{s})} |\bar{s}-s|$, i.e., $2 d$
is the minimal distance between $\bar{s}$ and ${\mathcal S}\backslash U(\bar{s})$.
For $n$ large enough, $|\phi_{t_n}(x)-\bar{s}| < d$;
hence the time $\tau$ it takes the flow to transport $\phi_{t_n}(x)$ out of $U(\bar{s})$
can be estimated by $\tau \geq d/F$. (We show below that the flow must indeed
transport $\phi_{t_n}(x)$ to ${\mathcal S}\backslash U(\bar{s})$ again, at least
for infinitely many $n$.)
Since $\dot{Z} > \varepsilon$ during the time $\tau$ we obtain
\begin{equation}
Z(\phi_{t_{n+1}}(x)) \geq Z(\phi_{t_n +\tau}(x)) \geq Z(\phi_{t_n}(x)) + \varepsilon\, d/F \ .
\end{equation}
for an infinite number of values of $n$.
This is a contradiction to the finiteness of $\lim_{n\rightarrow\infty} Z(\phi_{t_n}(x))$.

It remains to show that the flow $\phi_t$ transports $\phi_{t_n}(x)$ out of
a (sufficiently small) neighborhood $U(\bar{s})$ of $\bar{s}$ at least for infinitely many $n$.
Assume, indirectly, that the statement were false. Then, for all neighborhoods $V(\bar{s})$
there exists some $m\in\mathbb{N}$ such that
$\phi_{t_m+t}(x) \in V(\bar{s})$ for all $t \geq 0$, which means that
$\phi_{t}(x) \rightarrow \bar{s}$ ($t\rightarrow\infty$). Consequently,
since $\lim_{t\rightarrow\infty} Z(\phi_t(x))$ is finite,
the sequence $\dot{Z}(\phi_t(x))$ must contain a subsequence that converges to zero.
This is a contradiction to our assumption $\primelim_{x\rightarrow \bar{s}}\dot{Z}(x) \not\ni 0$.
\end{proof}

\end{appendix}

\end{document}